\documentstyle[epsfig]{elsart}

\begin{document}
\begin{frontmatter}
\title{Time Evolution of Decay Spectrum in $K^0, \, \overline{K^0} \rightarrow
 \pi^+ \pi^- e^+ e^-$}

\author{L.~M.~Sehgal and J.~van~Leusen}
\address{Institute of Theoretical Physics, RWTH Aachen, D-52056 Aachen, Germany}

\begin{abstract}
We consider the decay $K^0 \, (\overline{K^0}) \rightarrow \pi^+ \pi^- e^+ e^-$ of a 
neutral $K$ meson prepared in a state of strangeness $+1 \, (-1)$. The time evolution 
of the state produces remarkable time-dependent effects in the angular distribution
of the $\pi^+ \pi^- e^+ e^-$ system. These effects are correlated with the
time-dependence of the photon polarization in the radiative decay
$K^0 \, (\overline{K^0}) \rightarrow \pi^+ \pi^- \gamma$. We study, in particular, the 
$CP$-odd, $T$-odd term in the distribution $d\Gamma / d\phi$ of the angle between
the $\pi^+ \pi^-$ and the $e^+ e^-$ planes. We also give the spectrum in the case that
the decaying meson is an incoherent mixture of $K^0$ and $\overline{K^0}$, and discuss
the case of $K_S$ regeneration in a $K_L$ beam.
\end{abstract}
\end{frontmatter}

\section{Introduction}
The KTeV experiment~\cite{KTeVcoll} has measured a large $CP$-violating, $T$-odd 
asymmetry in the decay $K_L \rightarrow \pi^+ \pi^- e^+ e^-$, in quantitative agreement
with a prediction made some years ago~\cite{Sehgal:Wanninger,Heiliger:Sehgal}.
The origin of this effect lies in the amplitude of the radiative decay
$K_L \rightarrow \pi^+ \pi^- \gamma$, which contains a bremsstrahlung term proportional
to $\eta_{+-}$, as well as a $CP$-conserving direct emission term of magnetic dipole
character~\cite{E731coll}. The interference of the odd electric multipoles
$E1,\,E3,\,E5\, \cdots$ present in the bremsstrahlung amplitude, which all have
$CP = +1$,with the magnetic $M1$ multipole of $CP = -1$, produces $CP$-violating
components in the polarization state (Stokes vector) of the photon~\cite{Sehgal:Leusen}. The Dalitz
pair process $K_L \rightarrow \pi^+ \pi^- e^+ e^-$ acts as an analyser of the photon
polarization, exposing the $CP$-odd, $T$-odd component of the Stokes vector. The 
specific distribution that reveals the $CP$-violation is
\begin{equation}
\frac{d\Gamma}{d\phi} \sim 1 - \left(\Sigma_3 \cos 2\phi + \Sigma_1 \sin 2\phi \right)
\label{distKgamma}
\end{equation}
where $\phi$ is the angle between the $\pi^+ \pi^-$ and $e^+ e^-$ planes. The last term
in Eq. (\ref{distKgamma}) is $CP$-odd and $T$-odd, and produces an asymmetry
\begin{equation}
{\cal A}_{\phi} = \frac{\left( \int_{0}^{\pi/2} - \int_{\pi/2}^{\pi} +
\int_{\pi}^{3\pi/2} - \int_{3\pi/2}^{2\pi} \right) \frac{d\Gamma}{d\phi} d\phi}
{\left( \int_{0}^{\pi/2} + \int_{\pi/2}^{\pi} +
\int_{\pi}^{3\pi/2} + \int_{3\pi/2}^{2\pi}\right) \frac{d\Gamma}{d\phi} d\phi}
= - \frac{2}{\pi} \Sigma_1.
\end{equation}
The measured value~\cite{KTeVcoll} $\left| {\cal A}_{\phi} \right| = (13.6 \pm 2.5 
\pm 1.2) \%$ is in excellent agreement with the
prediction~\cite{Sehgal:Wanninger,Heiliger:Sehgal} of $14 \%$.

In a recent report~\cite{NA48coll}, the NA48 collaboration, while confirming the large 
$CP$-violating effect in $K_L \rightarrow \pi^+ \pi^- e^+ e^-$, has also studied the
decay $K_S \rightarrow \pi^+ \pi^- e^+ e^-$, finding no asymmetry in this case. This 
is entirely consistent with the fact that the amplitude 
$K_S \rightarrow \pi^+ \pi^- \gamma$ is accurately reproduced by bremsstrahlung alone,
so that no electric-magnetic interference is expected.

An interesting question raised by the above observations is the following: How does
the asymmetry ${\cal A}_{\phi}$ evolve with time if the neutral $K$ meson is prepared
in an initial state $K^0$ or $\overline{K^0}$ of definite strangeness? How does the 
value of ${\cal A}_{\phi}$ evolve from zero at short times (when the state decays
like $K_S$) to the value $- 14 \%$ at large times (when the state is essentially $K_L$)?
Finally, what type of evolution is expected in the decay of an incoherent 
$K^0 - \overline{K^0}$ mixture, such as an untagged neutral kaon beam originating in the 
decay $\phi \rightarrow K^0\overline{K^0}$?

This paper answers these questions by analysing in detail the time dependence of the 
decays $K^0, \, \overline{K^0} \rightarrow \pi^+ \pi^- e^+ e^-$. We study the evolution
of the full decay spectrum, especially the terms that are odd under $CP$. A comparison
is made between the behaviour of beams that are initially $K^0$, $\overline{K^0}$ or an
incoherent (untagged) mixture.

As a prelude to the discussion of the channel 
$K^0, \, \overline{K^0} \rightarrow  \pi^+ \pi^- e^+ e^-$, we analyse in 
Section~\ref{timevgamma} the time-dependence of the photon polarization in
$K^0, \, \overline{K^0} \rightarrow \pi^+ \pi^- \gamma$. This will reveal the behaviour
of the $CP$-odd components of the photon Stokes vector, one of which is reflected in
the asymmetry ${\cal A}_{\phi}$ in 
$K^0, \, \overline{K^0} \rightarrow \pi^+ \pi^- e^+ e^-$, that we discuss in 
Section~\ref{timevee}.

\section{Time Evolution of Photon Polarization in $K^0, \, \overline{K^0} \rightarrow
 \pi^+ \pi^- \gamma$}
\label{timevgamma}

The measured branching ratios and photon energy spectra in the decays
$K_L \rightarrow \pi^+ \pi^- \gamma$ and $K_S \rightarrow \pi^+ \pi^- \gamma$~\cite{E731coll}
can be well-described by the matrix elements~\cite{Sehgal:Wanninger}
\begin{eqnarray}
{\cal M} (K_S \rightarrow \pi^+ \pi^- \gamma) & = & e f_S
\left[\frac{\epsilon \cdot p_+}{k \cdot p_+} -
\frac{\epsilon \cdot p_-}{k \cdot p_-} \right] \nonumber\\
{\cal M} (K_L \rightarrow \pi^+ \pi^- \gamma) & = & e f_L
\left[\frac{\epsilon \cdot p_+}{k \cdot p_+} -
\frac{\epsilon \cdot p_-}{k \cdot p_-} \right] + e \frac{f_{DE}}{{M_K}^4}
\epsilon_{\mu\nu\rho\sigma} \epsilon^{\mu}k^{\nu}{p_+}^{\rho}{p_-}^{\sigma}
\end{eqnarray}
where
\begin{eqnarray}
& f_S = |f_S| e^{i \delta_0(s={M_K}^2)}, & \, f_L = \eta_{+-} f_S, \nonumber\\
& f_{DE} = |f_S| g_{M1}, & g_{M1} = i(0.76)e^{i \delta_1(s)}.
\nonumber
\end{eqnarray}
Here $\delta_{0,1}$ are the $\pi^+ \pi^-$ phase shifts in the $I=0$ $s$-wave and $I=1$
$p$-wave channel, respectively. Introducing the notation~\cite{Sehgal:Leusen}
\begin{eqnarray}
{\cal M}(K_{S,L} \rightarrow \pi^+ \pi^- \gamma) & = & 
\frac{e |f_S|}{{M_K}^4}  \left\{ E_{S,L}(\omega,\cos \theta) 
\left[\epsilon \cdot p_+ \, k \cdot p_- - \epsilon \cdot p_- \, k \cdot p_+ \right] 
\right. \nonumber\\
& & \left. \mbox{} + M_{S,L}(\omega, \cos \theta)
\epsilon_{\mu\nu\rho\sigma}\epsilon^{\mu}k^{\nu}{p_+}^{\rho}{p_-}^{\sigma}
\right\}
\end{eqnarray}
we have for the electric and magnetic amplitudes
\begin{eqnarray}
E_S = \left( \frac{2M_K}{\omega} \right)^2 
\frac{e^{i \delta_0(s={M_K}^2)}}{1-\beta^2 \cos^2 \theta}\, , & M_S & = 0\nonumber\\
E_L = \left( \frac{2M_K}{\omega} \right)^2 
\frac{\eta_{+-}e^{i \delta_0(s={M_K}^2)}}{1-\beta^2 \cos^2 \theta}\, , &
 M_L & = i(0.76)e^{i \delta_1(s)}. \label{defEandM}
\end{eqnarray}
As in~\cite{Sehgal:Leusen}, $\omega$ is the photon energy in the kaon rest frame, and 
$\theta$ is the angle of the $\pi^+$ relative to the photon in the $\pi^+ \pi^-$ 
c.m. frame. Noting that the strangeness eigenstates $K^0$ and $\overline{K^0}$ can be
written as
\begin{eqnarray}
K^0 & = & \left( K_S + K_L \right) / {\cal N} \nonumber\\
\overline{K^0} & = & \left( K_S - K_L \right) / \overline{{\cal N}}, \label{defK0K0bar}
\end{eqnarray}
the decay amplitudes for $K^0$ and $\overline{K^0}$ at time $t$ are
\begin{eqnarray}
{\cal M}(K^0(t) \rightarrow \pi^+ \pi^- \gamma) & \sim & 
\left\{ E(t,\omega,\cos \theta)
\left[\epsilon \cdot p_+ \, k \cdot p_- - \epsilon \cdot p_- \, k \cdot p_+ \right] 
\right. \nonumber\\
& & \left. \mbox{} + M(t, \omega, \cos \theta)
\epsilon_{\mu\nu\rho\sigma}\epsilon^{\mu}k^{\nu}{p_+}^{\rho}{p_-}^{\sigma}
\right\}\nonumber\\
{\cal M}(\overline{K^0}(t) \rightarrow \pi^+ \pi^- \gamma) & \sim & 
\left\{ \overline{E}(t,\omega,\cos \theta)
\left[\epsilon \cdot p_+ \, k \cdot p_- - \epsilon \cdot p_- \, k \cdot p_+ \right] 
\right. \nonumber\\
& & \left. \mbox{} + \overline{M}(t, \omega, \cos \theta)
\epsilon_{\mu\nu\rho\sigma}\epsilon^{\mu}k^{\nu}{p_+}^{\rho}{p_-}^{\sigma}
\right\}
\end{eqnarray}
where
\begin{eqnarray}
E & = & e^{- i \lambda_S t}E_S(\omega, \cos \theta) + e^{-i \lambda_L t}
E_L(\omega, \cos \theta) \nonumber\\
M & = & e^{-i \lambda_L t} M_L(\omega, \cos \theta) \nonumber\\
\overline{E} & = & e^{- i \lambda_S t}E_S(\omega, \cos \theta) - e^{-i \lambda_L t}
E_L(\omega, \cos \theta) \nonumber\\
\overline{M} & = & - e^{-i \lambda_L t} M_L(\omega, \cos \theta)
\label{amK0K0bar}
\end{eqnarray}
with $\lambda_{S,L} = m_{S,L} - \frac{i}{2} \Gamma_{S,L}$, $t$ being the proper time.

The amplitudes (\ref{amK0K0bar}) determine the Stokes vector of the photon in
$K^0, \, \overline{K^0} \rightarrow \pi^+ \pi^- \gamma$. For $K^0$ decay, the 
components of the Stokes vector $\vec{S} = \left(S_1, S_2, S_3 \right)$ at a time $t$ are
\begin{eqnarray}
S_1(t) & = & \frac{2\, {\rm Re} \left[ E^*(t)M(t) \right]}{|E(t)|^2 + |M(t)|^2} \nonumber\\
S_2(t) & = & \frac{2\, {\rm Im} \left[ E^*(t)M(t) \right]}{|E(t)|^2 + |M(t)|^2} \\
S_3(t) & = & \frac{|E(t)|^2 - |M(t)|^2}{|E(t)|^2 + |M(t)|^2}. \nonumber
\end{eqnarray}
The corresponding components for $\overline{K^0}$ decay are obtained by replacing
$E(t) \to \overline{E}(t)$, $M(t) \to \overline{M}(t)$.

The physical significance of $S_2(t)$ is that it represents the net circular polarization
of the photon in $K^0 \rightarrow \pi^+ \pi^- \gamma$ at time $t$:
\begin{equation}
S_2(t) = \frac{d\Gamma(\lambda_{\gamma} = -1,t) - d\Gamma(\lambda_{\gamma} = +1, t)}{
d\Gamma(\lambda_{\gamma} = -1,t) + d\Gamma(\lambda_{\gamma} = +1, t)}.
\end{equation}
The parameters $S_1(t)$ and $S_3(t)$, on the other hand describe the dependence of the
decay rate on the orientation of the polarization vector $\vec{\epsilon}$ relative to 
$\vec{n}_{\pi}$, the unit vector normal to the $\pi^+ \pi^-$ plane:
\begin{equation}
\frac{d\Gamma(t)}{d\phi} \sim 1 - \left(S_3(t) \cos 2\phi + S_1(t) \sin 2\phi \right)
\end{equation}
where the coordinates are chosen such that
\begin{equation}
\vec{k} = (0,0,k),\, \vec{\epsilon} = (\cos \phi, \sin \phi, 0),\, \vec{n}_{\pi} = 
\frac{\vec{p}_+ \times \vec{p}_-}{|\vec{p}_+ \times \vec{p}_-|} = (1,0,0).
\end{equation}

\begin{figure}
\centerline{\epsfxsize=8.0cm \epsfysize=5.7cm \epsffile{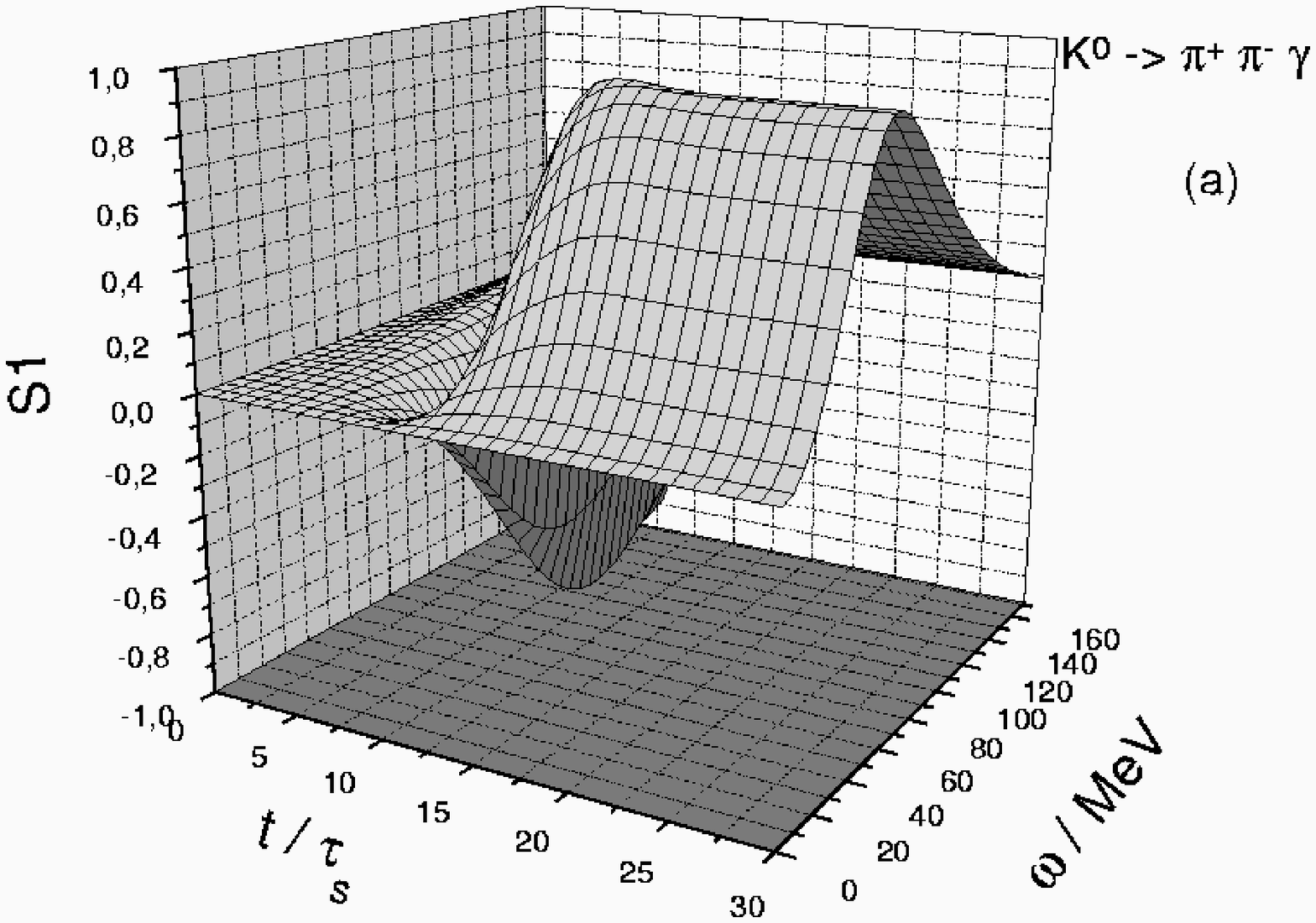}}
\centerline{\epsfxsize=8.0cm \epsfysize=5.7cm \epsffile{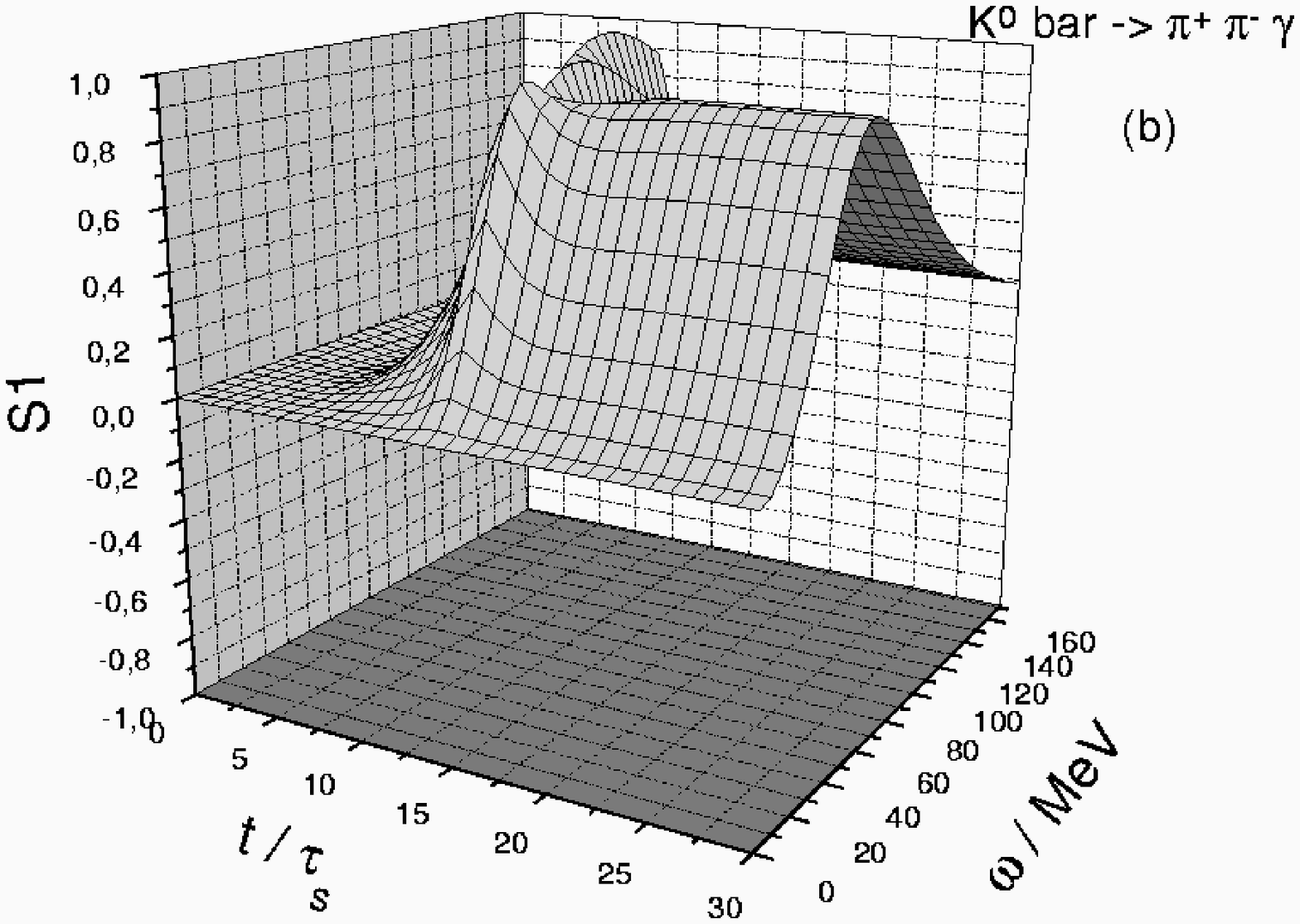}}
\centerline{\epsfxsize=8.0cm \epsfysize=5.7cm \epsffile{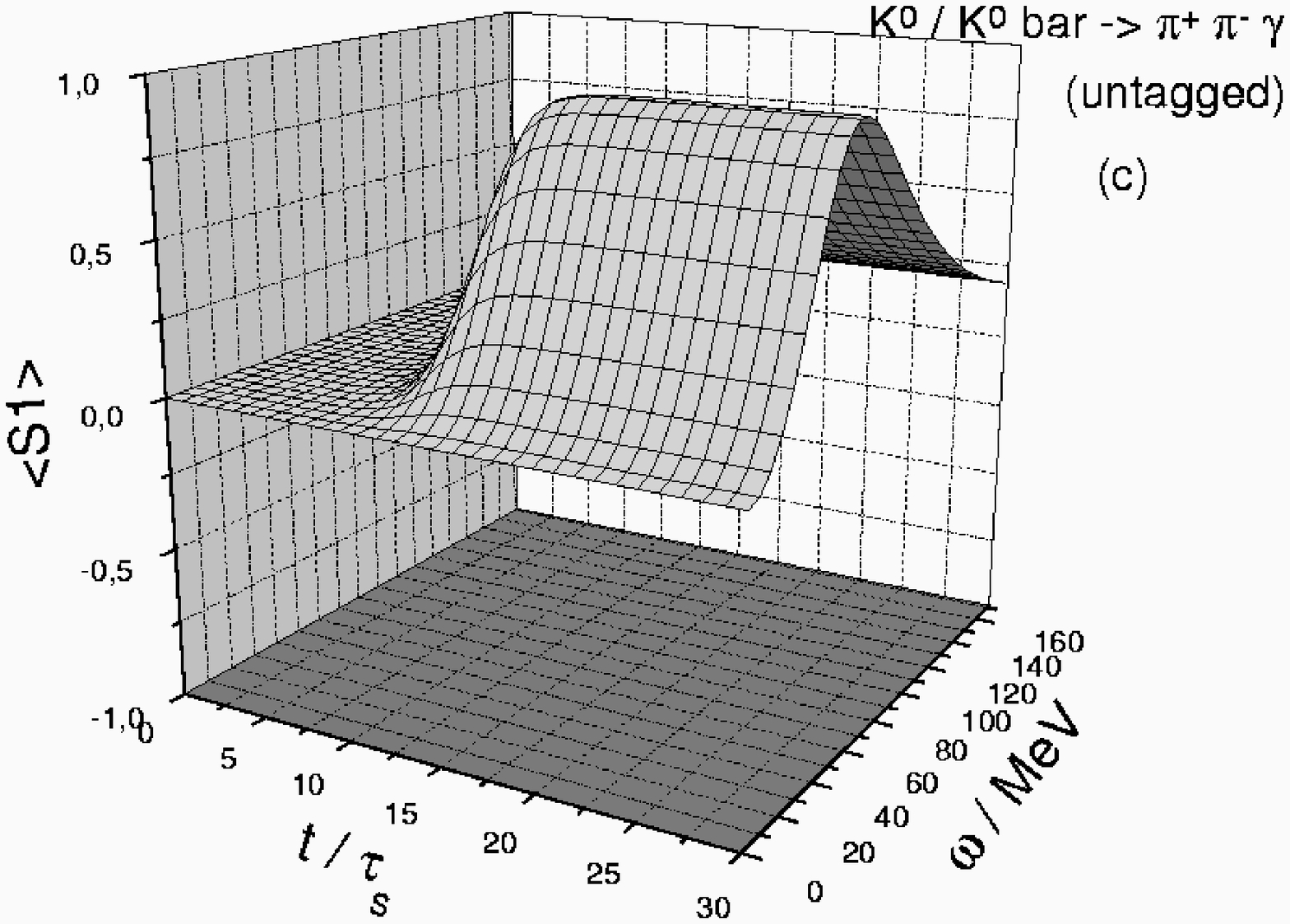}}
\caption{(a) Component $S1$ of the Stokes vector of the photon as a function of photon energy and time 
for the decay $K^0 \rightarrow \pi^+ \pi^- \gamma$; (b) same as (a) but for initial $\overline{K^0}$; (c)
 same as (a) but for an incoherent equal mixture of $K^0$ and $\overline{K^0}$.
\label{S13Dfig}
}
\end{figure}

\begin{figure}
\centerline{\epsfxsize=8.0cm \epsfysize=5.7cm \epsffile{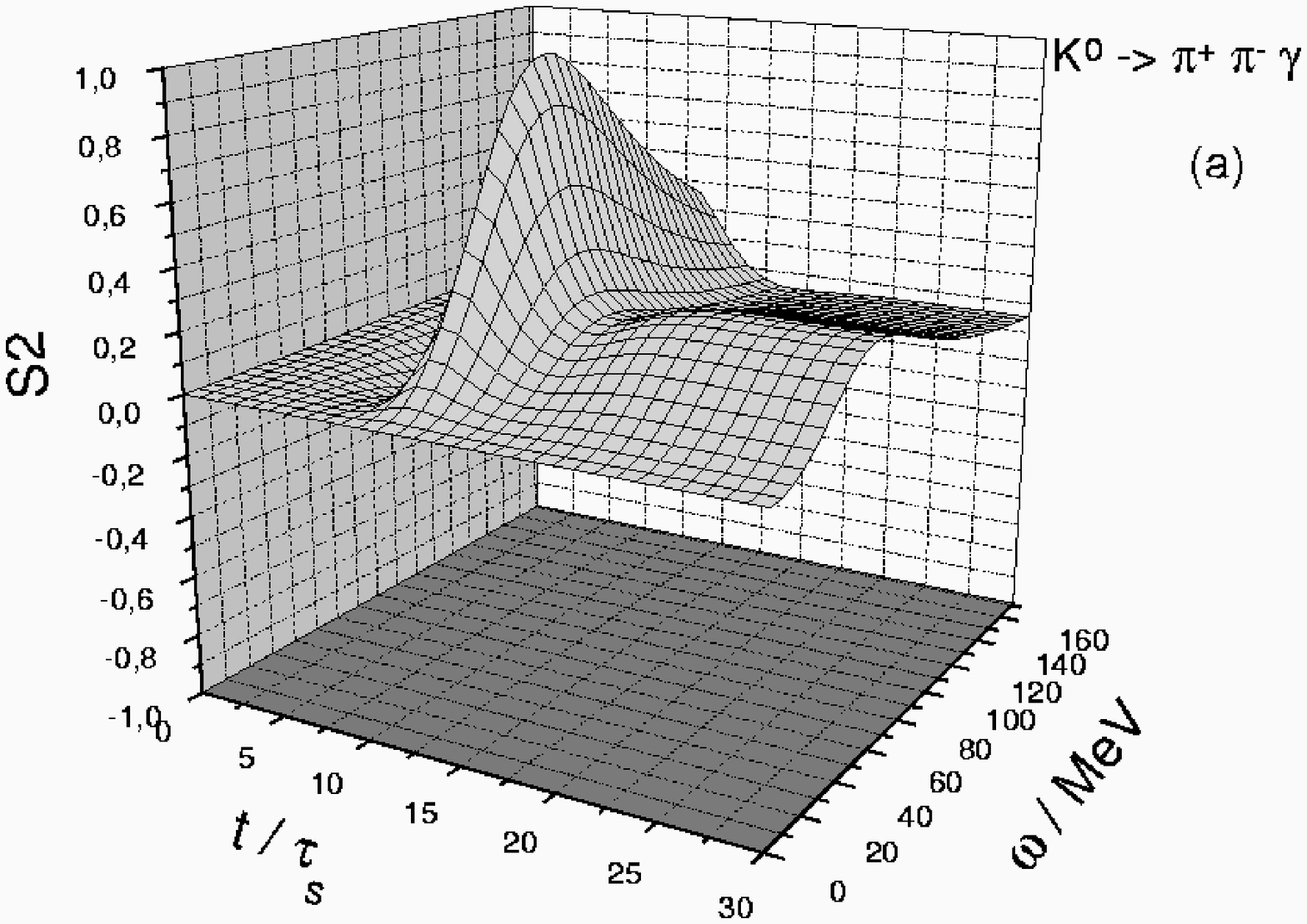}}
\centerline{\epsfxsize=8.0cm \epsfysize=5.7cm \epsffile{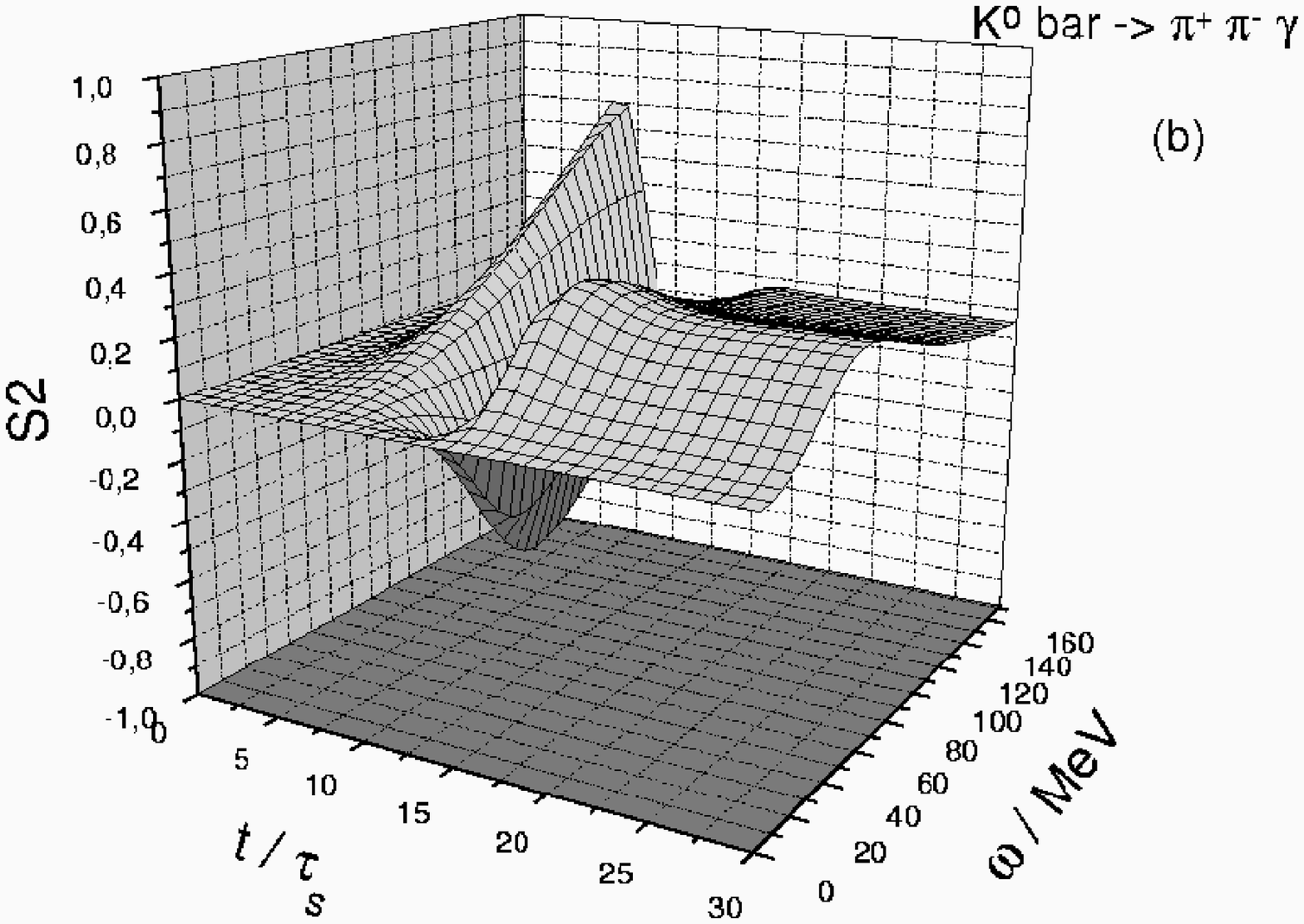}}
\centerline{\epsfxsize=8.0cm \epsfysize=5.7cm \epsffile{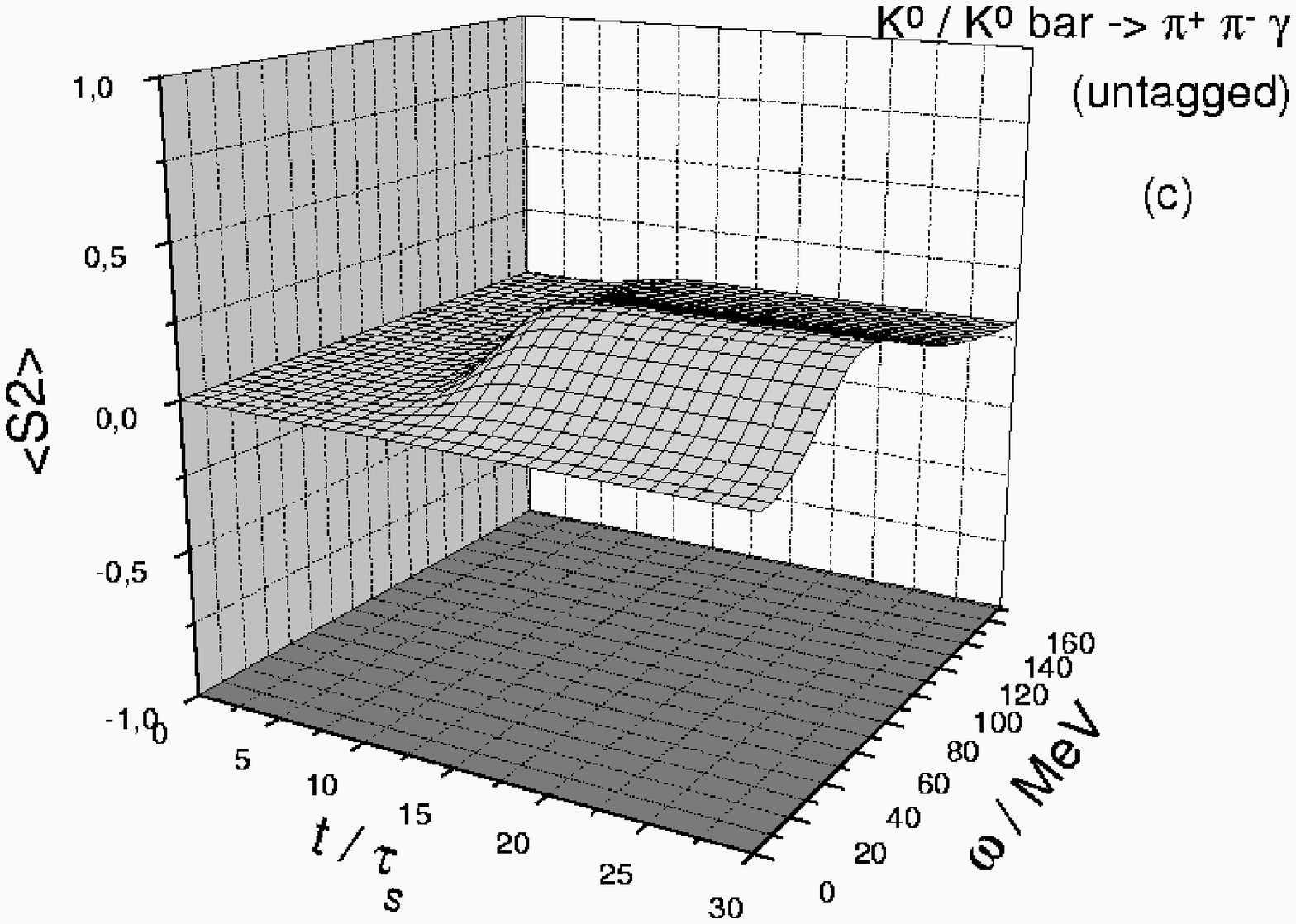}}
\caption{(a) Component $S2$ of the Stokes vector of the photon as a function of photon energy and time 
for the decay $K^0 \rightarrow \pi^+ \pi^- \gamma$; (b) same as (a) but for initial $\overline{K^0}$; (c)
 same as (a) but for an incoherent equal mixture of $K^0$ and $\overline{K^0}$.
\label{S23Dfig}
}
\end{figure}

In Figs.~\ref{S13Dfig}a and \ref{S23Dfig}a, we show the components $S_1(t)$ and $S_2(t)$
as functions of photon energy for the decay of an initial $K^0$. The corresponding
Stokes vector components for an initial $\overline{K^0}$ are shown in Figs.~\ref{S13Dfig}b 
and \ref{S23Dfig}b. Notice the intricate interference effect in the time-dependence, 
particularly in the region $t \sim 10 \tau_S$. Of special interest is the limiting case
$t \to \infty$, when the beam is essentially pure $K_L$. The Stokes vector components 
$S_1$ and $S_2$ then reduce to those shown in~\cite{Sehgal:Leusen} for the case 
$K_L \rightarrow \pi^+ \pi^- \gamma$. For comparison with the $K^0$ and $\overline{K^0}$
cases, we have also considered the case of an untagged initial beam, consisting of an 
incoherent equal mixture of $K^0$ and $\overline{K^0}$. The Stokes vector is then
\begin{eqnarray}
\left< S_1(t)\right> & = & \frac{2\, {\rm Re} 
\left[ E^*(t)M(t) + \overline{E}^*(t) \overline{M}(t) \right]}{
|E(t)|^2 + |M(t)|^2 +|\overline{E}(t)|^2 + |\overline{M}(t)|^2}\nonumber\\
\left< S_2(t)\right> & = & \frac{2\, {\rm Im} 
\left[ E^*(t)M(t) + \overline{E}^*(t) \overline{M}(t) \right]}{
|E(t)|^2 + |M(t)|^2 +|\overline{E}(t)|^2 + |\overline{M}(t)|^2} \\
\left< S_3(t) \right> & = & \frac{|E(t)|^2 - |M(t)|^2 +|\overline{E}(t)|^2 - |\overline{M}(t)|^2
}{|E(t)|^2 + |M(t)|^2 +|\overline{E}(t)|^2 + |\overline{M}(t)|^2}. \nonumber
\end{eqnarray}
The time-dependent parameters $\left< S_1(t) \right>$ and $\left< S_2(t) \right>$ are 
plotted in Figs.~\ref{S13Dfig}c and \ref{S23Dfig}c. Notice that the strong fluctuations
in $S_1(t)$ and $S_2(t)$ for the $K^0$ and $\overline{K^0}$ cases are now smoothed out.
For $t \gg \tau_S$, of course, all three cases give the same Stokes vector, namely that
corresponding to $K_L \rightarrow \pi^+ \pi^- \gamma$.

The polarization components $\left< S_1(t) \right>$ and $\left< S_2(t) \right>$ in the
incoherent case have a special significance: they represent $CP$-violating observables
{\it at any time $t$}. In the model under discussion, they vanish when $\eta_{+-} = 0$,
which is not the case for the components $S_1(t)$ and $S_2(t)$ originating from $K^0$ or
$\overline{K^0}$. Furthermore, in the ``hermitian limit''~\cite{Sehgal:Leusen} 
(i.e. $\delta_0 = \delta_1 = 0$,
${\rm arg} \, \eta_{+-} = \frac{\pi}{2}$) the parameter $\left< S_1(t) \right>$ survives, but
$\left< S_2(t) \right>$ vanishes. This is the hallmark that characterises the observable
$\left< S_1(t) \right>$ as being $CP$-odd, $T$-odd, and the observable $\left< S_2(t) \right>$
as being $CP$-odd, $T$-even~\cite{Sehgal}.

Of particular relevance in our discussion of 
$K^0, \, \overline{K^0} \rightarrow  \pi^+ \pi^- e^+ e^-$ is the behaviour of $S_1(t,\omega)$
portrayed in Fig.~\ref{S13Dfig} as a function of photon energy $\omega$. This will be
found to have a resemblance with the asymmetry ${\cal A}_{\phi}(t, s_{\pi})$ as a function
of the $\pi^+ \pi^-$ invariant mass (recall that for $K \rightarrow \pi^+ \pi^- \gamma$, 
$s_{\pi} = M^2_K - 2 M_K \omega$). We now proceed to a systematic analysis of time
dependence in the Dalitz pair reaction.

\section{Time Evolution of $K^0, \, \overline{K^0} \rightarrow 
\pi^+ \pi^- \lowercase{e}^+ \lowercase{e}^-$}
\label{timevee}

We begin by recalling the matrix element of the long-lived kaon decay 
$K_L \rightarrow \pi^+ \pi^- e^+ e^-$, treating the $e^+ e^-$ system as an internal conversion
pair associated with the radiative decay 
$K_L \rightarrow \pi^+ \pi^- \gamma$~\cite{Sehgal:Wanninger,Heiliger:Sehgal}. Writing the
matrix element in the form
\begin{eqnarray}
{\cal M}(K_L \rightarrow \pi^+ \pi^- e^+ e^-) & = & -2 \frac{G_F}{\sqrt{2}} \sin \theta_C
\left\{ \frac{1}{M_K} \left[ f (p_+ + p_-)_{\lambda} + g(p_+ - p_-)_{\lambda} \right] \right.
\nonumber\\
& & \left.\mbox{} + i \frac{h}{M_K^2} \, \epsilon_{\lambda \mu \nu \sigma} p_K^{\mu} 
(p_+ + p_-)^{\nu} (p_+ - p_-)^{\sigma} \right\} \nonumber\\
& & \times \mbox{} \overline{u}(k_-) \gamma^{\lambda} v(k_+),
\label{matKee}
\end{eqnarray}
the form factors $f$, $g$ and $h$ are related to the parameters $f_S$, $\eta_{+-}$ and $g_{M1}$
of the radiative decay as follows~\cite{Heiliger:Sehgal}:
\begin{eqnarray}
f & = & CM_K^4 |\eta_{+-}| e^{i(\delta(M^2_K) + \varphi_{+-})} \frac{1}{s_l} 
\frac{-4 \beta \cos \theta_{\pi}}{s(1-\beta^2 \cos^2 \theta_{\pi})} \nonumber\\
g & = & CM_K^4 |\eta_{+-}| e^{i(\delta(M^2_K) + \varphi_{+-})} \frac{1}{s_l}
\frac{4}{s(1-\beta^2 \cos^2 \theta_{\pi})}\\
h & = & -CM_K^4 \frac{1}{s_l}(0.76) e^{i \delta_1 (s_{\pi})} \nonumber
\end{eqnarray}
Here $\theta_{\pi}$ is the angle of the $\pi^+$ in the $\pi^+\pi^-$ rest frame, relative to
the sum of the $e^+$ and $e^-$ momenta in that frame; $s_l$ is the invariant mass of the 
lepton pair; $s_{\pi}$ is the $\pi^+ \pi^-$ invariant mass; and $s$ is defined as 
$s = \frac{1}{2} (M^2_K - s_{\pi} - s_l)$. The constant $C$ is given by
\begin{equation}
CM^4_K = \left( - \frac{G_F}{\sqrt{2}} \sin \theta_C \frac{1}{M_K} \right)^{-1} \pi \alpha |f_S|
\end{equation}

It is now easy to obtain from the amplitude ${\cal M}(K_L \rightarrow \pi^+ \pi^- e^+ e^-)$ 
the corresponding time-dependent amplitudes for $K^0$ and $\overline{K^0}$ decay by making
use of the following artifice: To obtain ${\cal M}(K^0(t) \rightarrow \pi^+ \pi^- e^+ e^-)$,
we make the replacement
\begin{eqnarray}
 f & \rightarrow & f \left( e^{-i\lambda_L t} + \frac{1}{\eta_{+-}} e^{-i \lambda_S t} \right) 
/ {\cal N} \nonumber\\
g & \rightarrow & g \left( e^{-i \lambda_L t}+ \frac{1}{\eta_{+-}} e^{-i \lambda_S t} \right) 
/ {\cal N} \label{repK0} \\
h & \rightarrow & h e^{-i \lambda_L t} / {\cal N} \nonumber
\end{eqnarray}
in Eq. (~\ref{matKee}). Similarly, to obtain 
${\cal M}(\overline{K^0}(t) \rightarrow \pi^+ \pi^- e^+ e^-)$, we make the replacement
\begin{eqnarray}
 f & \rightarrow & f \left( -e^{-i\lambda_L t} + \frac{1}{\eta_{+-}} e^{-i \lambda_S t} \right)
 / \overline{\cal N} \nonumber\\
g & \rightarrow & g \left( -e^{-i \lambda_L t}+ \frac{1}{\eta_{+-}} e^{-i \lambda_S t} \right) 
/ \overline{\cal N} \label{repK0bar} \\
h & \rightarrow & -h e^{-i \lambda_L t} / \overline{\cal N}. \nonumber
\end{eqnarray}
The normalization factors ${\cal N}$ and $\overline{\cal N}$ in Eqs.~(\ref{repK0})
and (\ref{repK0bar}) are those in the states $K^0$ and $\overline{K^0}$ 
(Eq.~(\ref{defK0K0bar})), where ${\cal N} = 2p$, $\overline{\cal N} = 2q$, with
$|p|^2 - |q|^2 = \frac{2 \, {\rm Re}\, \eta_{+-}}{1 + |\eta_{+-}|^2}$ and $|p|^2+|q|^2 = 1$. 
With this prescription in mind,
we will continue to use the formalism developed in~\cite{Heiliger:Sehgal} for
$K_L \rightarrow \pi^+ \pi^- e^+ e^-$, with the understanding that in dealing with $K^0$
and $\overline{K^0}$ decay, the form factors $f$, $g$ and $h$ are to be replaced by the
time-dependent combinations given in Eqs.~(\ref{repK0}) and (\ref{repK0bar}).

\begin{figure}
\centerline{\epsfxsize=8.0cm \epsfysize=5.7cm \epsffile{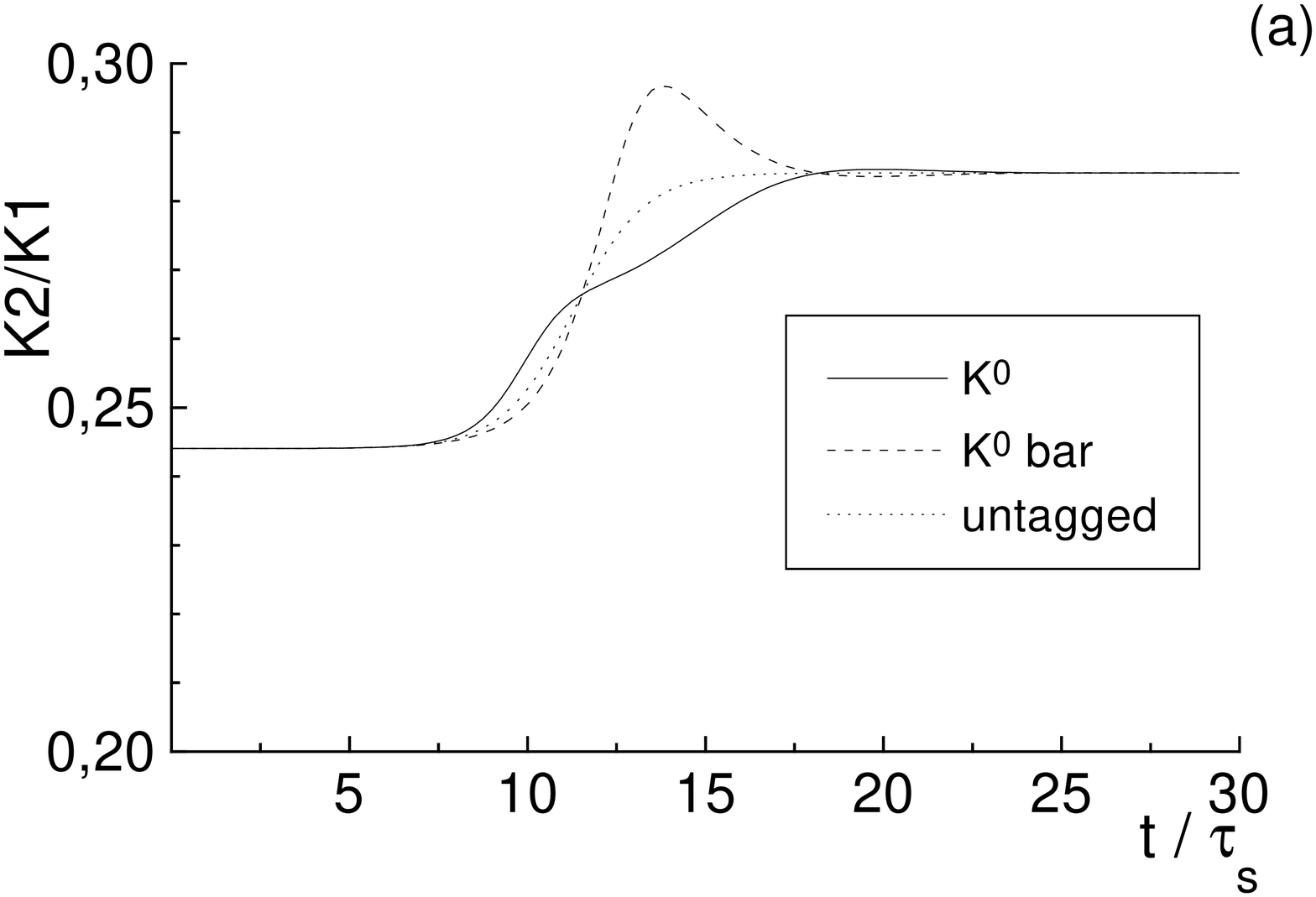}}
\centerline{\epsfxsize=8.0cm \epsfysize=5.7cm \epsffile{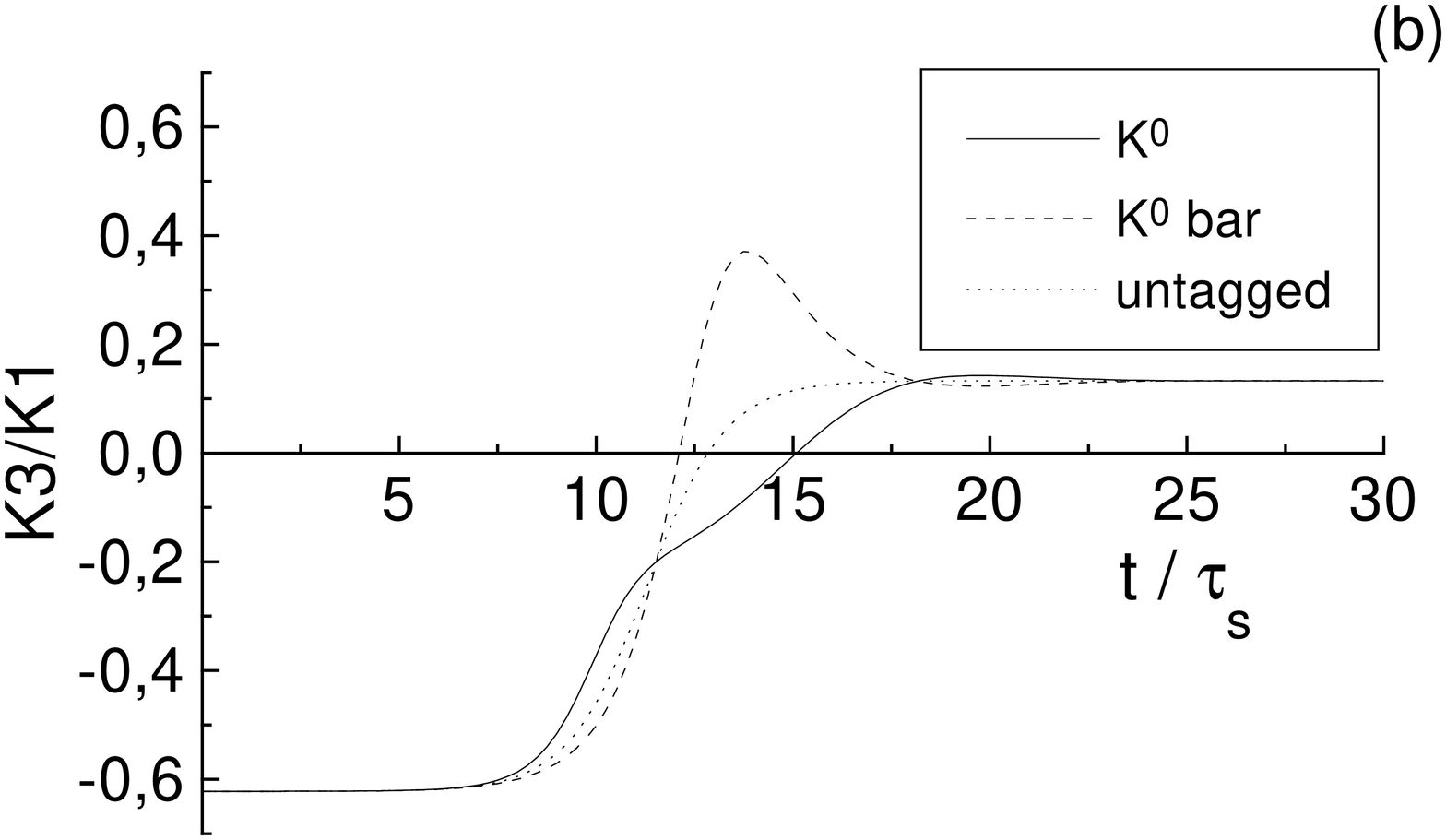}}
\centerline{\epsfxsize=8.0cm \epsfysize=5.7cm \epsffile{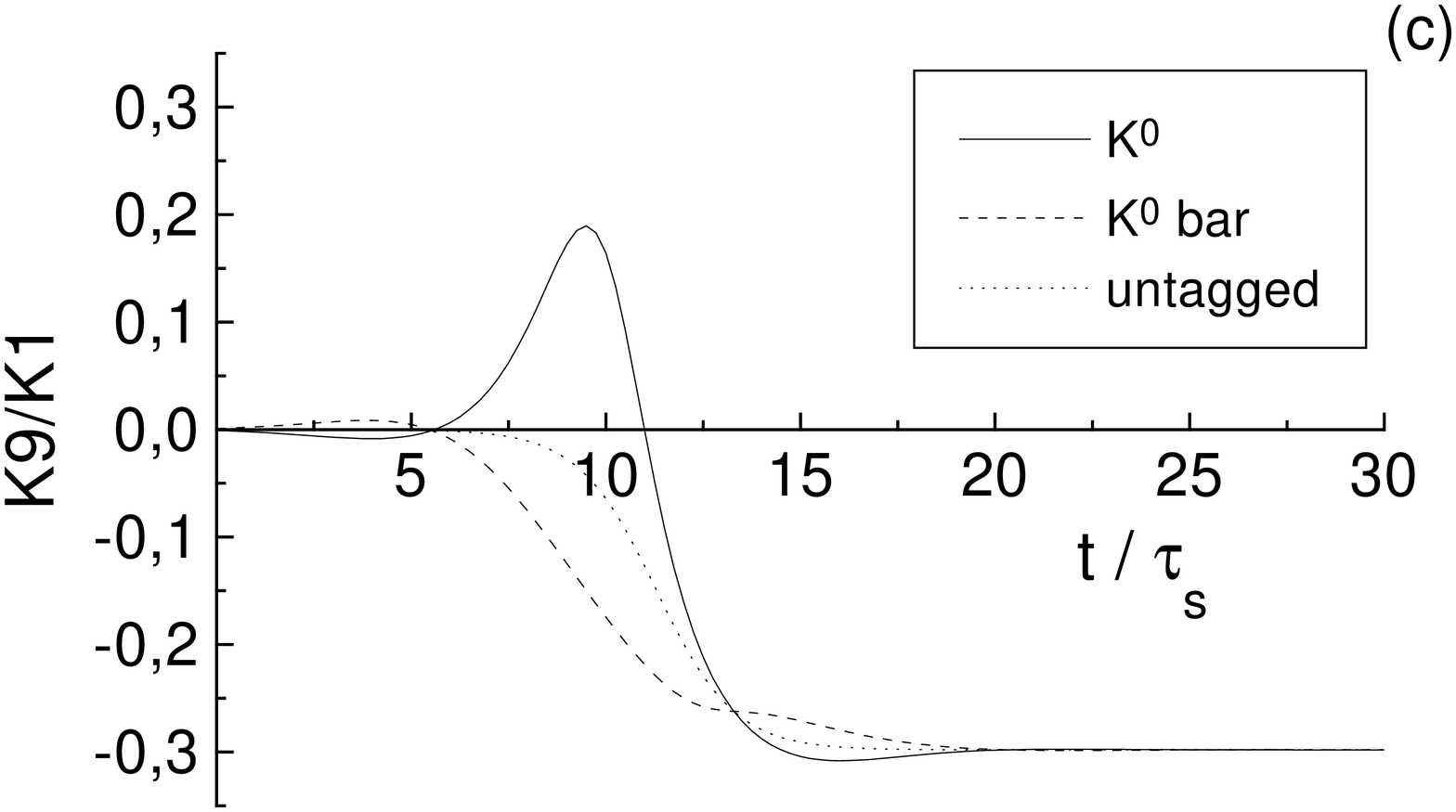}}
\caption{Time-dependent coefficients $K_i/K1$: (a) $K2/K1$ for the decays $K^0$ 
and $\overline{K^0} \rightarrow \pi^+ \pi^- e^+ e^-$ as well as for an untagged 
initial beam; (b) same as (a) but for $K3/K1$; (c) same as (a) but for $K9/K1$.
\label{KitoK1fig}
}
\end{figure}

Following the procedure of Ref.~\cite{Heiliger:Sehgal}, we have calculated the angular 
distribution of the decays $K^0(t) \rightarrow \pi^+ \pi^- e^+ e^-$ and
$\overline{K^0}(t) \rightarrow \pi^+ \pi^- e^+ e^-$ in the form
\begin{equation}
\frac{d\Gamma}{d \cos \theta_l \, d \phi} = K_1 + K_2 \cos 2 \theta_l + K_3 \sin^2 \theta_l
\cos 2 \phi + K_9 \sin^2 \theta_l \sin 2 \phi. \label{defKi}
\end{equation}
Here $\theta_l$ is the angle of $e^+$ relative to the dipion momentum vector in the $e^+ e^-$
frame; and $\phi$ is the angle between the $\pi^+ \pi^-$ and $e^+ e^-$ planes. The last term in
Eq.~(\ref{defKi}) is odd under the $CP$-transformation $\vec{p}_{\pm} \to - \vec{p}_{\mp}$,
$\vec{k}_{\pm} \to - \vec{k}_{\mp}$, as well as under the $T$-transformation
$\vec{p}_{\pm} \to - \vec{p}_{\pm}$, $\vec{k}_{\pm} \to - \vec{k}_{\pm}$.
The time-dependent coefficients $K_2 / K_1$, $K_3 / K_1$ and $K_9 / K_1$ are shown in 
Fig.~\ref{KitoK1fig}, where we compare the cases $K^0$ and $\overline{K^0}$, and also show
the result for an incoherent $K^0 - \overline{K^0}$ mixture. This figure depicts the manner
in which the coefficients of the angular distribution evolve to their asymptotic values
appropriate to the decay $K_L \rightarrow \pi^+ \pi^- e^+ e^-$, namely $K_2/K_1 = 0.297$,
$K_3/K_1 = 0.180$, $K_9/K_1 = - 0.309$~\cite{Heiliger:Sehgal}.

\begin{figure}
\centerline{\epsfxsize=8.0cm \epsfysize=5.7cm \epsffile{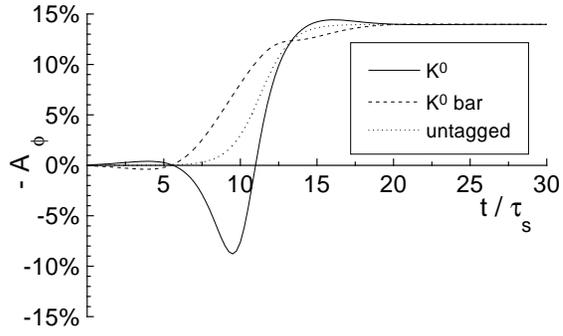}}
\caption{Time-dependent Asymmetry ${\cal A}_{\phi}$ for the decays $K^0$ and 
$\overline{K^0} \rightarrow \pi^+ \pi^- e^+ e^-$ as well as for an incoherent
equal mixture.
\label{asytfig}
}
\end{figure}

\begin{figure}
\centerline{\epsfxsize=8.0cm \epsfysize=5.7cm \epsffile{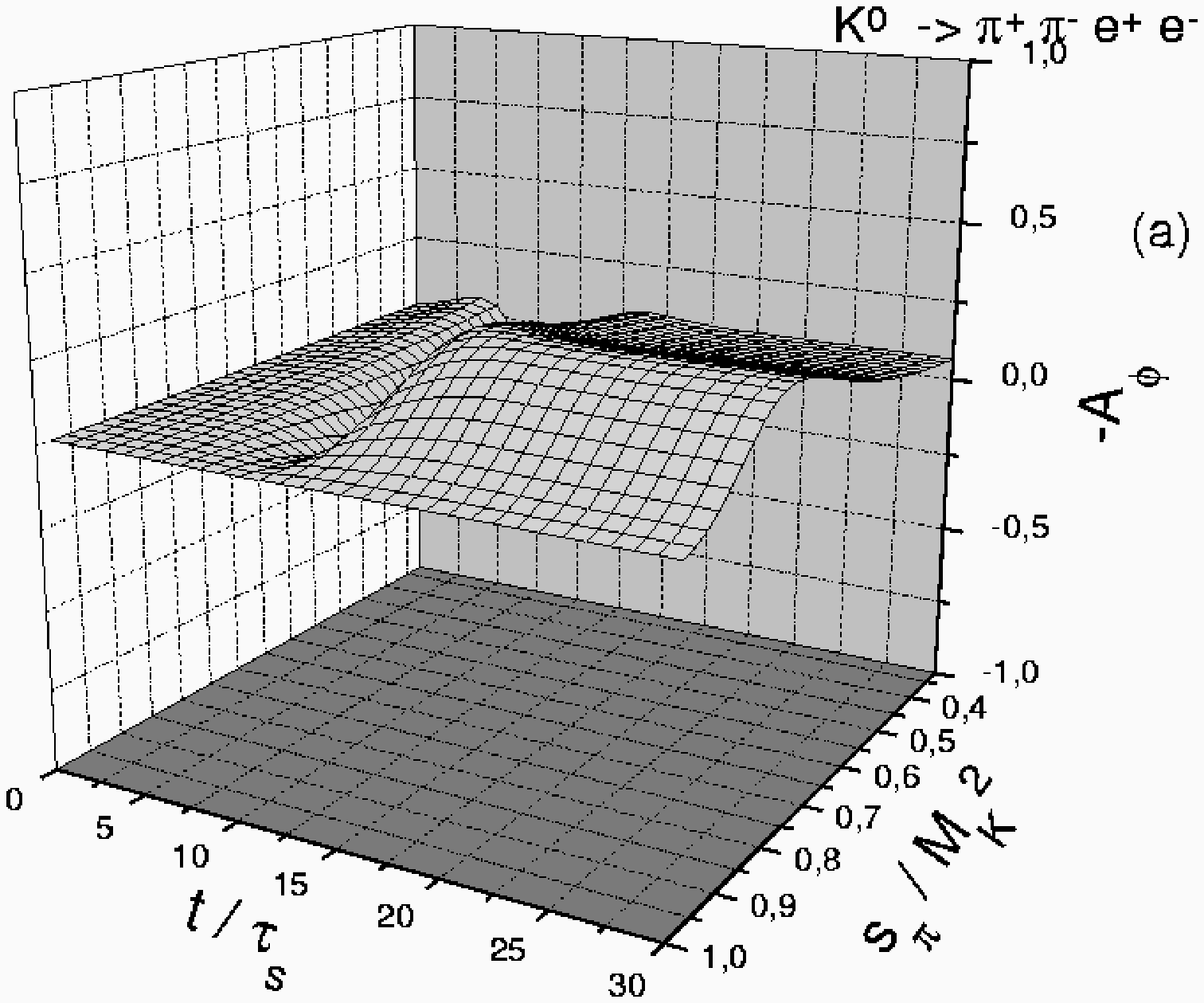}}
\centerline{\epsfxsize=8.0cm \epsfysize=5.7cm \epsffile{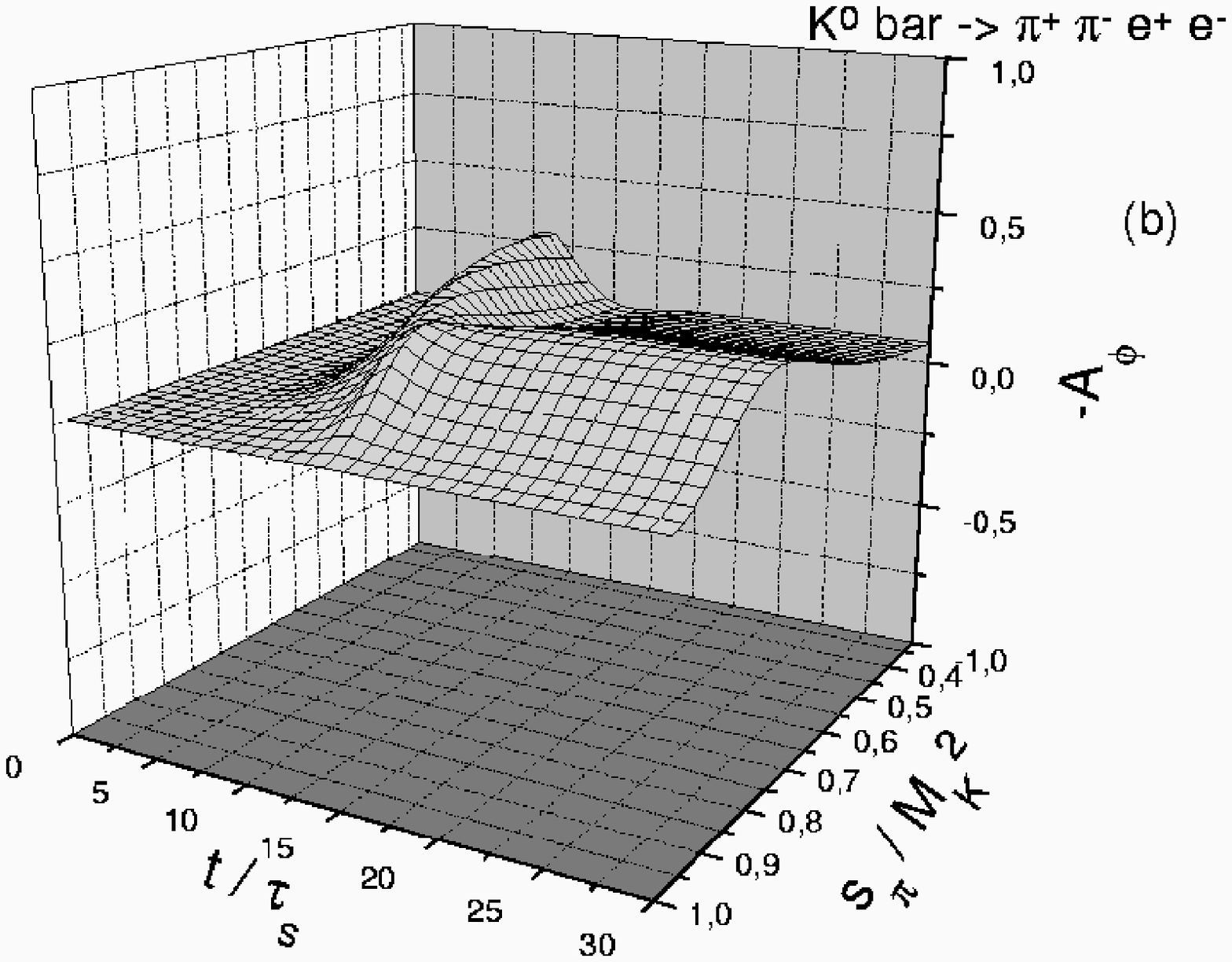}}
\centerline{\epsfxsize=8.0cm \epsfysize=5.7cm \epsffile{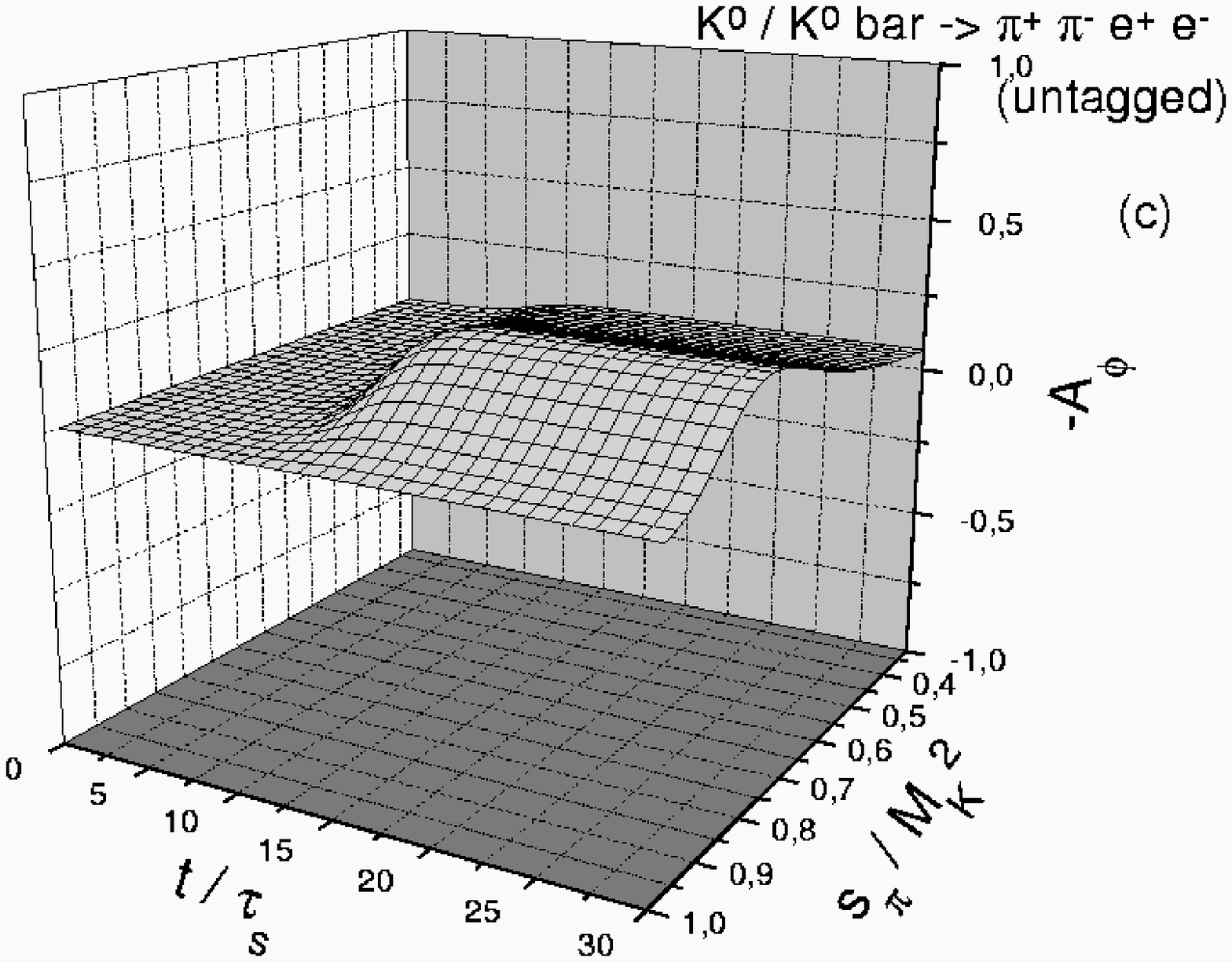}}
\caption{(a) The asymmetry ${\cal A}_{\phi}$ as a function of $s_{\pi}$ and time for the decay
$K^0 \rightarrow \pi^+ \pi^- e^+ e^-$; (b) same as (a) but for initial $\overline{K^0}$; 
(c) same as (a) but for an incoherent equal mixture of $K^0$ and $\overline{K^0}$.
\label{asy3Dfig}
}
\end{figure}

Integrating Eq.~(\ref{defKi}) over $\cos \theta_l$, we obtain the $\phi$-distribution
\begin{equation}
\frac{d \Gamma}{d \phi} \sim \left( 1 - \frac{1}{3} \frac{K_2}{K_1} \right)
+ \frac{2}{3} \left( \frac{K_3}{K_1} \cos 2 \phi + \frac{K_9}{K_1} \sin 2 \phi \right)
\end{equation}
which corresponds to an asymmetry
\begin{equation}
{\cal A}_{\phi} = 
\frac{2}{\pi} \frac{\frac{2}{3} \frac{K_9}{K_1}}{1-\frac{1}{3}\frac{K_2}{K_1}}
\end{equation}
The time-dependence of this asymmetry is exhibited in Fig.~\ref{asytfig}, which is the
answer to the question of how this asymmetry evolves from the value zero, appropriate to 
$K_S$ decay, to the value $-14\%$ observed for $K_L$ decay. We have also studied the 
time-dependent asymmetry ${\cal A}_{\phi}$ as a function of the $\pi^+ \pi^-$ invariant
mass. As seen in Fig.~\ref{asy3Dfig}, the function ${\cal A}_{\phi}(t, s_{\pi})$ has a
similarity to the Stokes vector $S_1(t,\omega)$ plotted in Fig.~\ref{S13Dfig}, confirming
the expectation that the $CP$-odd, $T$-odd term in the $\phi$-distribution of
$K^0 \rightarrow \pi^+ \pi^- e^+ e^-$ is correlated with the $CP$-odd, $T$-odd component
of the Stokes vector in $K^0 \rightarrow \pi^+ \pi^- \gamma$.

\begin{figure}
\centerline{\epsfxsize=8.0cm \epsfysize=5.7cm \epsffile{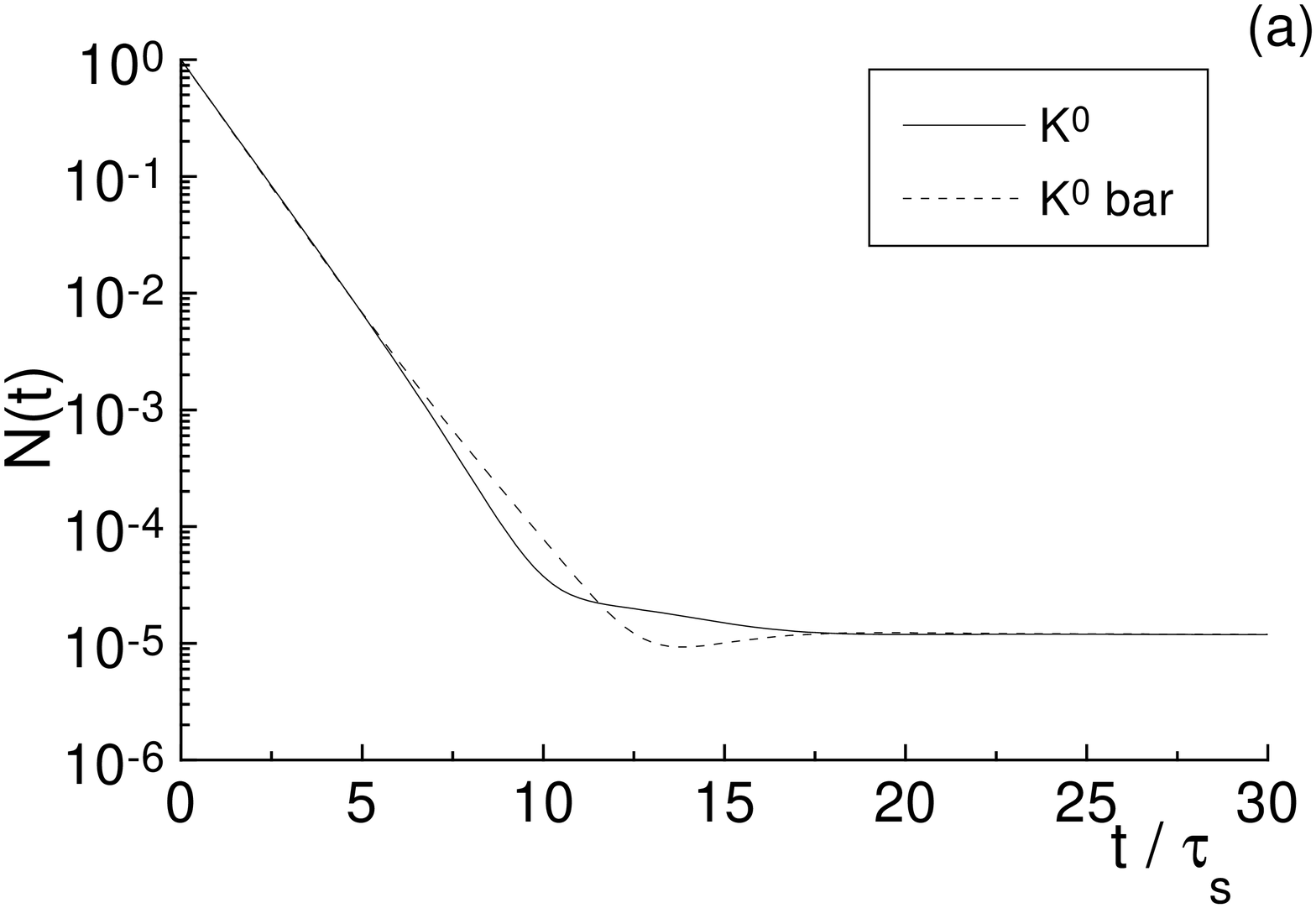}}
\centerline{\epsfxsize=8.0cm \epsfysize=5.7cm \epsffile{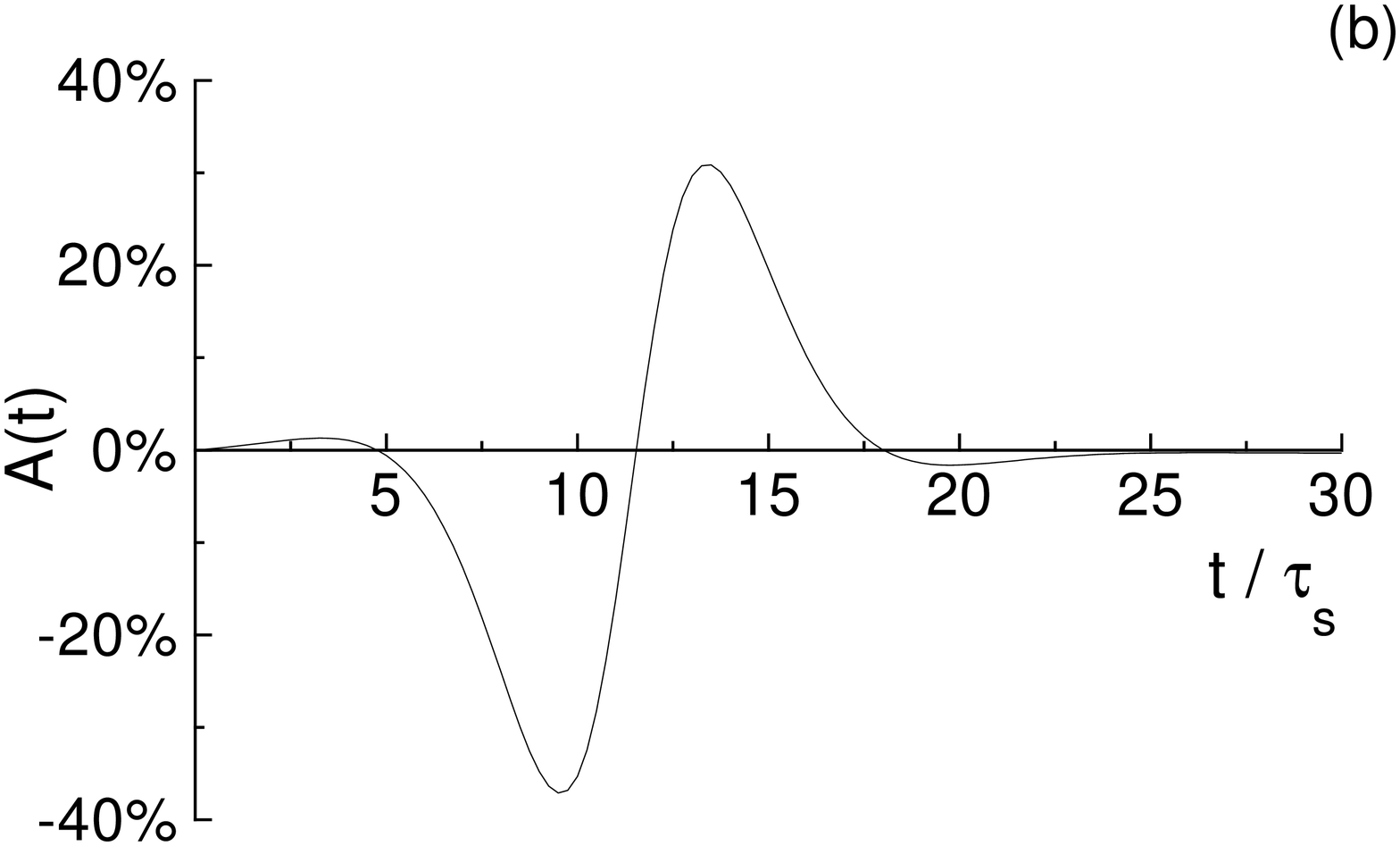}}
\caption{(a) Spectrum integrated decay rate for $K^0$ and 
$\overline{K^0} \rightarrow \pi^+ \pi^- e^+ e^-$; 
(b) asymmetry in the decay rate ${\cal A}(t) = \left( N(t)-\overline{N}(t) \right) /
\left( N(t)+\overline{N}(t) \right)$.
\label{specintfig}
}
\end{figure}

Finally, the spectrum-integrated decay rate in
$K^0 \rightarrow \pi^+ \pi^- e^+ e^-$ and $\overline{K^0} \rightarrow \pi^+ \pi^- e^+ e^-$
contains the standard $K_L-K_S$ interference effect proportional to $\eta_{+-}$, and a
corresponding asymmetry between $K^0$ and $\overline{K^0}$, which is shown in 
Fig.~\ref{specintfig}.

\section{Additional Remarks}

\begin{figure}
\centerline{\epsfxsize=8.0cm \epsfysize=5.7cm \epsffile{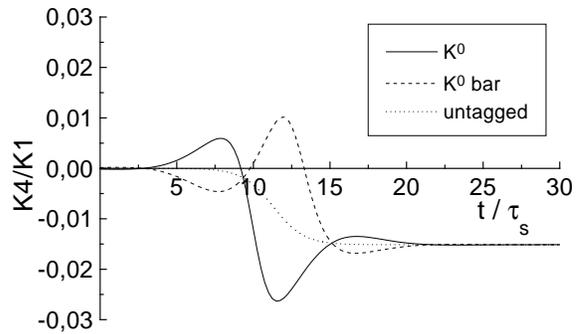}}
\caption{Coefficient $K4/K1$ generated by a charge-radius term for $K^0 \rightarrow \pi^+ \pi^- e^+ e^-$,
$\overline{K^0} \rightarrow \pi^+ \pi^- e^+ e^-$ and an untagged initial beam.
\label{K4toK1fig}
}
\end{figure}

(i) Our analysis of the time-dependence in 
$K^0, \, \overline{K^0} \rightarrow \pi^+ \pi^- e^+ e^-$ assumed the amplitude to be
entirely determined by the radiative decay
$K^0, \, \overline{K^0} \rightarrow \pi^+ \pi^- \gamma$. A non-radiative contribution to
the amplitude $K_L \rightarrow \pi^+ \pi^- e^+ e^-$ is possible in the form of a
``charge-radius'' term, with the characteristic feature of producing $\pi^+ \pi^-$ in an
$s$-wave. Such a configuration is not possible in the radiative decay 
$K_L \rightarrow \pi^+ \pi^- \gamma$. We have investigated the effects of a charge radius
term nominally parametrized by the coefficient $g_P = 0.15 e^{i \delta_1}$ defined 
in~\cite{Heiliger:Sehgal}. It has the interesting consequence of inducing a small term
of the form $K_4 \sin 2 \theta_l \cos \phi$ in the angular distribution 
$d\Gamma / d \cos \theta_l \, d\phi$ given in Eq.~(\ref{defKi}). Such a term is $CP$-odd
but $T$-even~\cite{Heiliger:Sehgal,Sehgal}. In Fig.~\ref{K4toK1fig}, we show the magnitude 
and time-dependence of the
coefficient $K_4/K_1$ generated by a charge-radius term $g_P$ with the nominal value given
above.

\begin{figure}
\centerline{\epsfxsize=8.0cm \epsfysize=5.7cm \epsffile{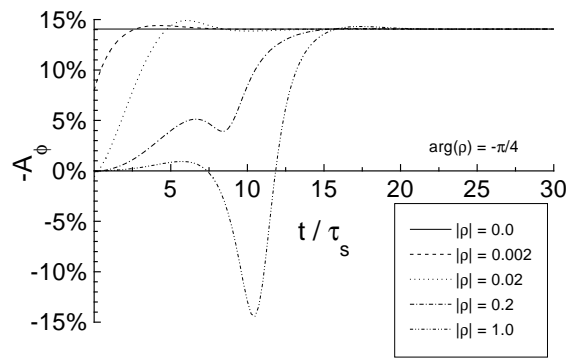}}
\caption{Time-dependent asymmetry ${\cal A}_{\phi}$ for a regenerated beam $K_L + \rho K_S$.
\label{asyregfig}
}
\end{figure}

(ii) Since the time-dependent interference effects in the decay spectrum of
$K^0, \, \overline{K^0} \rightarrow \pi^+ \pi^- e^+ e^-$ are strongest in the region
$t \sim 10 \tau_s$, one can ask whether similar effects could be observed with a beam of
the form $K_L + \rho K_S$, where $\rho$ is a regeneration amplitude induced by passage
of $K_L$ through material. We have calculated the decay spectrum of 
$(K_L + \rho K_S) \rightarrow \pi^+ \pi^- e^+ e^-$ as a function of time, for some typical
regeneration parameters $\rho = |\rho|e^{i\varphi_{\rho}}$, such as 
$|\rho| = 0.002$, $0.02$ and $0.2$ with $\varphi_{\rho} = - \pi/4$. In particular,
the time-dependent asymmetry ${\cal A}_{\phi}$ for such a beam is shown in 
Fig.~\ref{asyregfig}. It is clear that a suitable choice of a regenerator would permit a
study of the $CP$-odd, $T$-odd feature in the $\pi^+\pi^-e^+e^-$ decay of a neutral
K meson, as well as other aspects of the decay spectrum.

(iii) Several refinements in the matrix element of 
$K^0, \, \overline{K^0} \rightarrow \pi^+ \pi^- e^+ e^-$ can be imagined. There is a small
short-distance contribution associated with the interaction 
$s\overline{d} \rightarrow e^+ e^-$~\cite{Heiliger:Sehgal}, which, however, is highly 
suppressed on account of the
small quark mixing factor $V_{ts}V^*_{td}$. The magnetic dipole term in the radiative
amplitude $K_L \rightarrow \pi^+\pi^-\gamma$, represented by the coupling constant 
$g_{M1}$, actually has a certain dependence on $s_{\pi}$~\cite{Cox:Belz}, which can be 
incorporated into the analysis. Finally, the radiative amplitudes 
$K_{L,S} \rightarrow \pi^+ \pi^- \gamma$ could contain additional components such as a direct
$E1$ multipole in $K_S$ decay or a direct $E2$ term in $K_L$ decay (see, e.g.~\cite{Ambrosio:Isidori}).
 The former would give 
rise to a departure from pure bremsstrahlung in the photon energy spectrum of 
$K_S \rightarrow \pi^+ \pi^- \gamma$, while the latter would produce a $CP$-violating
charge asymmetry in the Dalitz plot of $K_L \rightarrow \pi^+ \pi^- \gamma$. There is also
the possibility of a direct $CP$-violating $E1$ multipole in $K_L$ decay, that would show up
as a difference between $\eta_{+-\gamma}$ and $\eta_{+-}$ in the time-dependence of 
$K_L + \rho K_S \rightarrow \pi^+ \pi^- \gamma$. 

A specific consequence of introducing direct $E1$ multipoles in $K_L \rightarrow \pi^+ \pi^- \gamma$
is to modify the amplitude $E_L$ in Eq.(\ref{defEandM}) to
\begin{equation}
E_L = \left( \frac{2 M_K}{\omega} \right)^2 \frac{\eta_{+-} e^{i \delta_0}}{1-\beta^2 \cos^2 \theta}
+ g_{E1} e^{i(\varphi_{+-} + \delta_1)} + i \tilde{g}_{E1} e^{i \delta_1}
\end{equation}
where $g_{E1}$ is a measure of direct ($CP$-conserving) $E1$ emission in
$K_S \rightarrow \pi^+ \pi^- \gamma$, and $\tilde{g}_{E1}$ a measure of a direct
$CP$-violating $E1$ emission in $K_L \rightarrow \pi^+ \pi^- \gamma$ 
($g_{E1}$ and $\tilde{g}_{E1}$ being real). The resulting change in the
asymmetry ${\cal A}_{\phi}$ in $K_L \rightarrow \pi^+ \pi^- e^+ e^-$ is~\cite{Sehgal:Wanninger}
\begin{equation}
{\cal A}_{\phi} = 15\% \sin \left( \varphi_{+-} + \delta_0 - \delta_1 \right) + 38 \% \left[
\frac{g_{E1}}{|g_{M1}|} \sin \varphi_{+-} + \frac{\tilde{g}_{E1}}{|g_{M1}|} 
\right].
\end{equation}
The observed branching ratio for $K_S \rightarrow \pi^+ \pi^- \gamma$ limits $|g_{E1}/g_{M1}|$ to
$< 5\%$~\cite{Sehgal:Wanninger}, while typical estimates of $|\tilde{g}_{E1}/g_{M1}|$ are of
order  $10^{-3}$~\cite{Ambrosio:Isidori}.

Needless
to say, any incisive study of the channels $K^0, \, \overline{K^0} \rightarrow \pi^+ \pi^- e^+ e^-$ 
and $K_{L,S} \rightarrow \pi^+\pi^-\gamma$ should be alert to the possible
presence of correction terms of the form described above.

\end{document}